%% file: 00-main.tex
\documentclass[sigconf]{acmart}
\pdfoutput=1

\usepackage{array}
\newcolumntype{L}{>{\centering\arraybackslash}m{8.5cm}}
\usepackage{booktabs} 

\setcopyright{rightsretained}

\acmDOI{10.475/123_4}

\acmISBN{123-4567-24-567/08/06}

\acmConference[KDD'18]{ACM SIGKDD Conference on Knowledge Discovery and Data Mining}{August 2018}{London, UK}
\acmYear{2018}
\copyrightyear{2018}

\acmPrice{15.00}

\usepackage{epstopdf}
\usepackage{multirow}
\usepackage{balance}
\usepackage{amsfonts}
\usepackage{amsmath}
\usepackage{amssymb}
\usepackage{graphicx}
\usepackage{subfigure}
\usepackage[noend,ruled]{algorithm2e}
\usepackage{microtype}
\usepackage{url}
\usepackage{enumitem}
\usepackage{wrapfig,lipsum}

\newtheorem{thm:def}{Definition}
\newtheorem{thm:eg}{Example}
\newtheorem{thm:lem}{Lemma}
\newtheorem{thm:obs}{Observation}
\newtheorem{thm:req}{Requirement}

\newcommand{\nop}[1]{}

\newcommand{\ie}{{\sl i.e.}}

\newcommand{\seed}{\mathbf{seeds}}
\newcommand{\core}{\mathbf{c}}
\newcommand{\g}{\mathbf{g}}

\DeclareMathOperator*{\argmax}{argmax}

\DeclareMathAlphabet{\mathbbold}{U}{bbold}{m}{n}

\newcommand{\smallsection}[1]{\vspace{1mm}\noindent\textbf{#1.}}    

\begin{document}

\title{Contrast Subgraph Mining from Coherent Cores}
\author{Jingbo Shang$^1$, Xiyao Shi$^1$, Meng Jiang$^2$, Liyuan Liu$^1$, Timothy Hanratty$^3$, Jiawei Han$^1$}
\affiliation{%
  \institution{$^1$Department of Computer Science, University of Illinois Urbana-Champaign, IL, USA}
}
\affiliation{%
  \institution{$^2$Department of Computer Science and Engineering, University of Notre Dame, IN, USA}
}
\affiliation{%
  \institution{$^3$US Army Research Laboratory, MD, USA}
}

\affiliation{%
  \institution{$^1$\{shang7, xshi27, ll2, hanj\}@illinois.edu $\quad$ $^2$mjiang2@nd.edu $\quad$ $^3$timothy.p.hanratty.civ@mail.mil}
}

\renewcommand{\shortauthors}{J. Shang et al.}

\begin{abstract}
	\input{0-abstract}
\end{abstract}

\maketitle

\input{1-introduction}
\input{3-preliminary}

\input{2-related_work}

\input{4-method}
\input{5-experiment}
\input{6-conclusion}

\bibliographystyle{abbrv}
\bibliography{cited}

\end{document}

%% file: 0-abstract.tex

Graph pattern mining methods can extract informative and useful patterns from large-scale graphs and capture underlying principles through the overwhelmed information.
Contrast analysis serves as a keystone in various fields and has demonstrated its effectiveness in mining valuable information.
However, it has been long overlooked in graph pattern mining.
Therefore, in this paper, we introduce the concept of contrast subgraph, 
that is, a subset of nodes that have significantly different edges or edge weights in two given graphs of the same node set.
The major challenge comes from the gap between the contrast and the informativeness.
Because of the widely existing noise edges in real-world graphs, the contrast may lead to subgraphs of pure noise. 
To avoid such meaningless subgraphs, we leverage the similarity as the cornerstone of the contrast. 
Specifically, we first identify a coherent core, which is a small subset of nodes with similar edge structures in the two graphs, and then induce contrast subgraphs from the coherent cores. 
Moreover, we design a general family of coherence and contrast metrics and derive a polynomial-time algorithm to efficiently extract contrast subgraphs. 
Extensive experiments verify the necessity of introducing coherent cores as well as the effectiveness and efficiency of our algorithm.
Real-world applications demonstrate the tremendous potentials of contrast subgraph mining. 

%% file: 1-introduction.tex

\section{Introduction}

Graph pattern mining methods can extract useful information from large-scale graphs and capture underlying principles through the overwhelmed information. 
Such information has benefited many fields, such as community analysis, document retrieval, and human mobility analysis.

Recently, graph patterns across multi-layer/multi-view graphs, or graphs having different edge structures of the same node set, have drawn lots of attentions. 
For example, frequent subgraph mining aims to discover the similar patterns in different graphs~\cite{yan2002gspan,yan2003closegraph}. 
Many researchers tried to identify the subsets of nodes that form dense subgraphs in most of the graphs~\cite{hu2005mining,pei2005mining,jiang2009mining} or identify discriminative (sometimes also named as ``contrast''~\cite{ting2006mining}) structures from frequent subgraphs using node labels~\cite{ting2006mining,thoma2010discriminative,jin2011lts}.

Contrast analysis serves as a keystone in many data mining problems, while the related study in graph pattern mining has been long overlooked.
For example, in the fields of text mining and document retrieval, given a major corpus and a background corpus of the same vocabulary, much effort has been paid to identify informative phrases whose frequencies in the major corpus are significantly higher than those in the background corpus~\cite{monroe2008fightin,bedathur2010interesting,gao2012top,majumdar2014fast,liu2015mining,shang2017automated}. 
Meanwhile, very few attempt has been made to contrast subgraph mining. 
Analogous to document comparative analysis, if we compare two graphs, $G_A$ and $G_B$, of the same set of nodes $V$, we should be able to find some informative subgraph $\g$ whose edge weights in $G_A$ are significantly different from those in $G_B$. 
In this paper, we name such subgraph $\g$ as a \emph{contrast subgraph}.

\begin{figure}[t]
  \centering
  \includegraphics[width=0.5\textwidth]{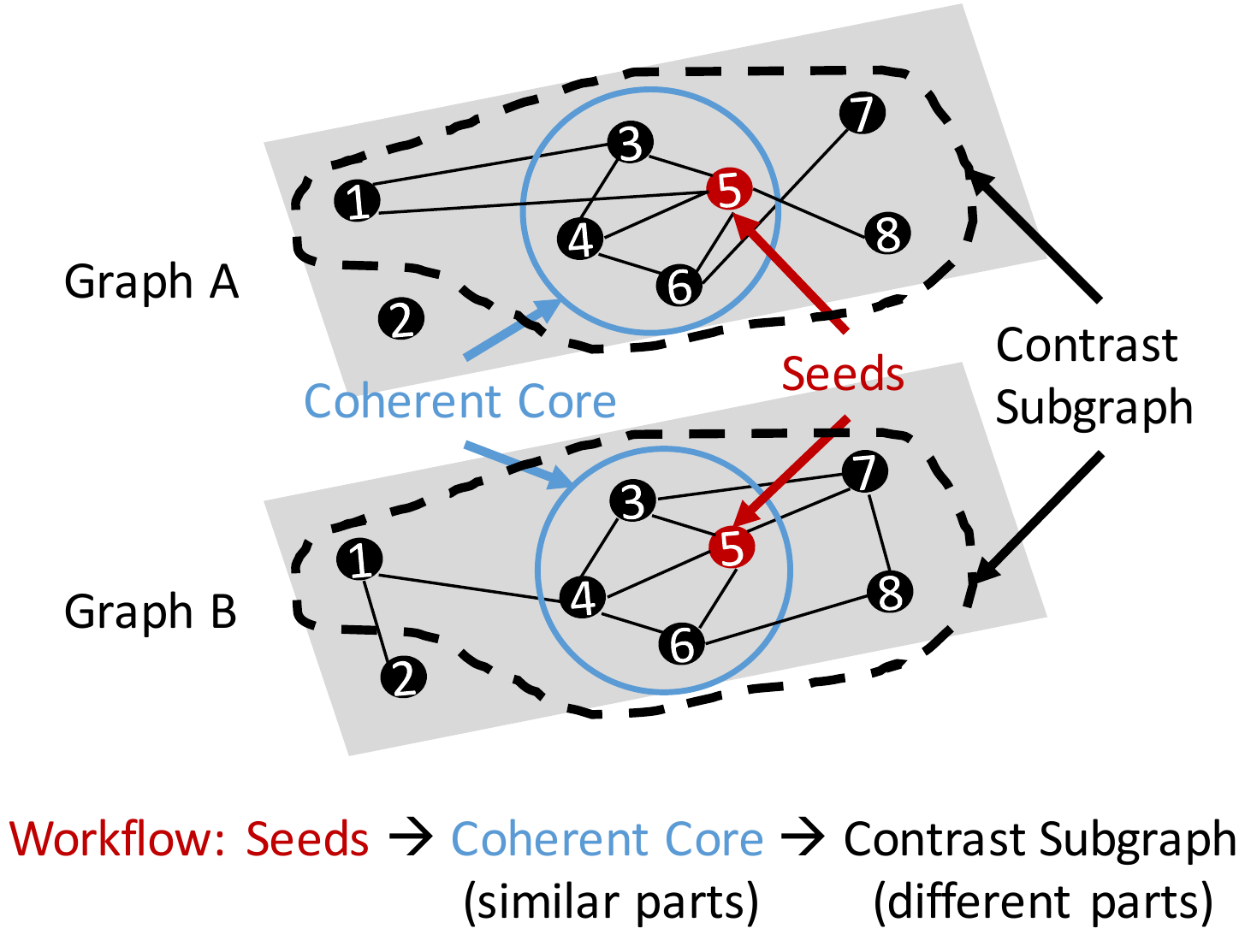}
  \vspace{-0.3cm}
  \caption{Workflow visualization using hypothetical graphs.}\label{fig:overview}
  \vspace{-0.3cm}
\end{figure}

\begin{figure*}[t]
  \centering
  \includegraphics[width=1.0\textwidth]{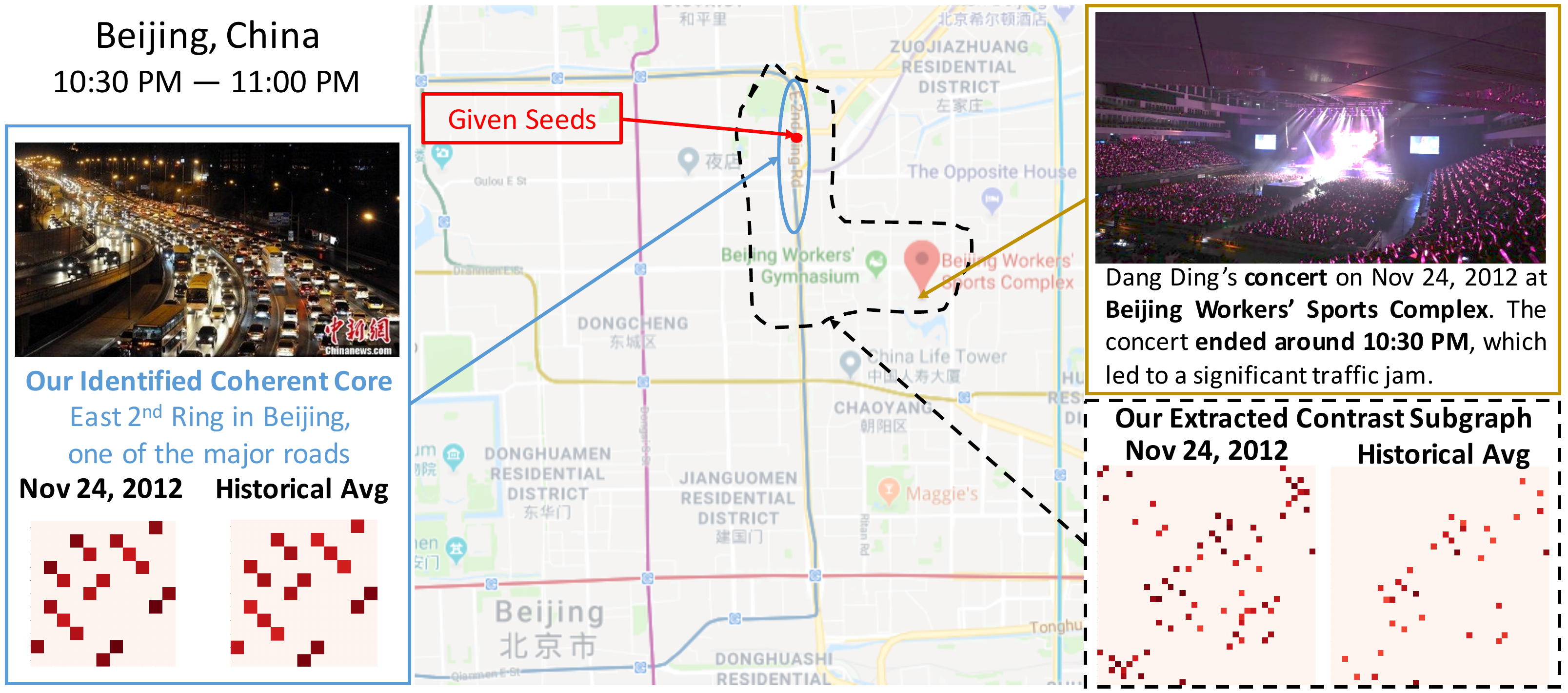}
  \vspace{-0.3cm}
  \caption{Spatio-Temporal Event Detection. 
  Heat maps visualize the adjacency matrices within the corresponding regions.
  The darker, the more traffic.
  Contrasting two road networks of the real-time traffic and historical averages,
  the identified coherent core demonstrates the usual, busy traffic on the East 2nd Ring in Beijing,
  while the extracted contrast subgraph reveals an unusually large volume of traffic, which indicates the ending of a concert event at the Beijing Workers' Sports Complex. }\label{fig:contrast_concert}
  \vspace{-0.3cm}
\end{figure*}

Contrast subgraphs have great potentials to facilitate many downstream applications such as temporal role identification~\cite{mccallum2005topic,wang2010mining,rossi2015role,henderson2012rolx}, online-offline community detection~\cite{wellman2002networked,kavanaugh2005community,zhang2011integrating,buote2009exploring,wellman2002networked}, and spatio-tempral event detection~\cite{hong2015detecting,zhang2016geoburst}. 
For example, contrast subgraphs can provide useful signals for advisor-advisee identification tasks~\cite{wang2010mining}. 
Moreover, when both online and offline social networks are provided, mining contrast subgraphs can detect some unusual relationships~\cite{buote2009exploring,wellman2002networked}. 
Also, by contrasting the incoming and outgoing real-time traffics, one can detect specific types of spatio-temporal events~\cite{hong2015detecting}.

Besides extracting contrast subgraphs, defining the contrast itself remains an open problem.
Moreover, trying to induce \emph{informative} subgraphs from the contrast makes the task even more challenging.
In real-world graphs, there usually exist noise edges.
For the algorithms directly optimizing the contrast, it is likely to end up with a subset of nodes which have heavy noise in one of the graphs and no edge in the other graph. 
To avoid such numerous but meaningless subgraphs, we propose to induce contrast subgraphs from the \emph{coherent cores}, i.e., a subset of nodes with similar edge structures in $G_A$ and $G_B$.
We visualize the workflow using two hypothetical graphs in Figure~\ref{fig:overview}.
Given the seed node $5$, we first identify the coherent core of nodes $\{3, 4, 5, 6\}$, and then add nodes $1$, $7$, and $8$ into the contrast graph.
More experiments in Section~\ref{exp:coherent_core} demonstrate that inducing from the coherent cores is necessary and it acts as an anchor to guarantee the informativeness of the induced contrast subgraphs.

Beyond the problem formulation, it is also challenging to develop an efficient and scalable algorithm for mining contrast subgraphs from large-scale graphs. 
A straightforward solution is to apply co-dense graph mining algorithms on $(G_A, \overline{G_B})$ and then on $(\overline{G_A}, G_B)$. 
However, the dense complementary graph (i.e., $\overline{G_A}$ or $\overline{G_B}$) makes most of, if not all, pruning techniques initially designed for sparse graphs ineffective. 
In this paper, we derive a polynomial-time algorithm to efficiently identify coherent subgraph cores and then extract contrast subgraphs. 
More specifically, we apply a binary search on the coherent/contrast score and construct a network such that whether the current score is achievable is equivalent to whether the min $\mathcal{S}-\mathcal{T}$ cut in the network is above a certain threshold. 
Thanks to the duality between the min cut and the max flow, we can solve the reduced problem in a polynomial time~\cite{orlin2013max}.

Our experimental results based on real-world datasets demonstrate the identified coherent core and the extracted contrast subgraph are quite insightful and encouraging in a variety of tasks.
For example, based on the traffic data and road network in Beijing, contrast subgraphs between the real-time traffic and the historical traffic indicate the events happening in the city.
Figure~\ref{fig:contrast_concert} shows the coherent core and the contrast subgraph discovered by our proposed method through a comparison between the traffic data on Nov 24, 2012, and the historical averages.
The usual, busy traffic on the major roads (i.e., the East 2nd Ring in Beijing) form the coherent core.
The contrast subgraph reveals the unusual traffic around the Beijing Workers' Sports Complex between 10:30~--~11:00~PM.
We find that the contrast subgraph indicates the ending of Dang Ding's concert event, which further reflects the precious value of the proposed contrast subgraph mining.

To our best knowledge, this is the first work that aims to mine informative contrast subgraphs between two graphs. 
Our contributions are highlighted as follows.
\begin{itemize}[leftmargin=*,noitemsep]
    \item We formulate the contrast subgraph mining problem and avoid meaningless contrast subgraphs by inducing from coherent cores. The problem is formulated for multiple applications using a general family of metrics.
    \item We derive a polynomial-time algorithm to efficiently extract contrast subgraphs from large-scale graphs.
    \item Extensive experiments verify the necessity of introducing coherent cores as well as the effectiveness and efficiency of our algorithm. Real-world applications demonstrate the tremendous potentials of contrast subgraph mining. 
\end{itemize}

\noindent \textbf{Reproducibility:} We release our code at the GitHub\footnote{\url{https://github.com/shangjingbo1226/ContrastSubgraphMining}}.

The remainder of this paper is organized as follows. 
We first formulate the contrast subgraph mining problem in Section~\ref{sec:pre}. 
The related work is discussed in Section~\ref{sec:rel}. 
Section~\ref{sec:method} covers the technical details of our derived polynomial-time algorithm.
In Section~\ref{sec:exp}, we use a real-world task to verify the necessity of inducing contrast subgraph from coherent cores and evaluate the efficiency of our algorithm.
Section~\ref{sec:app} uses two real-world applications to demonstrate the importance of contrast subgraph mining.
Section~\ref{sec:con} concludes this paper and outlines future directions.

%% file: 3-preliminary.tex

\section{Contrast Subgraph Mining}\label{sec:pre}

    In this section, we first formulate the problem and propose a general family of metrics. 
    And then, we discuss our choices of specific metrics used in our experiments.

    \subsection{Problem Formulation}
        The input of the contrast subgraph mining problem consists of \emph{two weighted, undirected} graphs $G_A = (V, E_A)$ and $G_B = (V, E_B)$.
        The node set $V$ is the same in both $G_A$ and $G_B$.
        But the edge weights in $E_A$ and $E_B$ are different.
        As summarized in Table~\ref{tbl:notation}, we use $E_A(u, v)$ and $E_B(u,v)$ to denote the edge weights of the edge between nodes $u$ and $v$ in these two graphs, respectively.
        Because $G_A$ and $G_B$ are undirected graphs, we have $\forall u,\ v$, $E_A(u ,v) = E_A(v, u)$ and $E_B(u ,v) = E_B(v, u)$.

        In this paper, without loss of generality, we make two assumptions for the input graphs.
        First, we assume $E_A(u, v)$, $E_B(u, v)$ $\ge 0$, where the edge weight $0$ means there is no edge.
        In most of the applications, the edge weights are mainly about the connection strengths between nodes.
        And we only care about the weight difference between the corresponding edges.
        Therefore, it is reasonable to assume the weights are non-negative.
        Note that this assumption includes boolean edges as a special case.
        Second, we assume the preprocessing has been conducted on $E_A$ and $E_B$, thus the ranges or distributions of $E_A(u, v)$ and $E_B(u, v)$ would be similar.

        The user can provide \emph{seed nodes}, if necessary, to better facilitate her interests.
        To make sure the nodes are really around the given seeds or the coherent core, we define the \emph{neighbor nodes} by a parameter $r$,
        \begin{equation}
            N_r(\mathbf{s}) = \{ u | d_A(\mathbf{s}, u) \le r \vee d_B(\mathbf{s}, u) \le r \}
        \end{equation}
        where $\mathbf{s}$ is a subset of nodes, $d_A(\mathbf{s}, u)$ and $d_B(\mathbf{s}, u)$ are the minimum number of non-zero edges to traverse from the node $u$ to any node in $\mathbf{s}$. 
        In practice, the value of $r$ can be chosen based on the diameter of the two input graphs.
        For example, the collaborator graph induced from the DBLP publication dataset has a small diameter, so $r=1$ and $r=2$ are good choices.
        The road network has a relatively large diameter, so we choose $r$ from $10$ to $20$.
        Reasonable values of $r$ can always lead to meaningful results.

        In summary, the inputs are two undirected, weighted graphs $G_A$ and $G_B$, a parameter $r$, and seed nodes $\seed$ (could be empty).
        As we discussed, to avoid meaningless cases, we propose to induce contrast subgraphs from the \emph{coherent cores}, where a coherent core is a subset of nodes with similar edges in the two graphs.
        The goal is to first identify the coherent core $\core$ around the seed nodes $\seed$ ($\seed \subset \core \subset N_r(\seed)$), and then induce the contrast subgraph $\g$ from this coherent core ($\core \subset \g \subset N_r(\core)$).

    \begin{table}[t]
        \center
        \caption{Notation Table.}
        \vspace{-0.3cm}
        \label{tbl:notation}
    \scalebox{0.85}{
        \begin{tabular}{|c|l|}
        \hline
        $G_A, G_B$ & Two undirected, weighted input graphs. \\
        \hline
        $\overline{G_A}, \overline{G_B}$ & Complementary graphs considering whether the edge exists. \\
        \hline
        $V$ & The node set for both $G_A$ and $G_B$. \\
        \hline
        $E_A(u, v), E_B(u, v)$ & Edge weights between nodes $u$ and $v$ in $G_A$ and $G_B$. \\
        \hline
        $\seed$ & A set of seed nodes. It could be empty. \\
        \hline
        $\core$ & Coherent core. A subset of $V$. $\seed \subset \core$.\\
        \hline
        $\g$ & Contrast subgraph. A subset of $V$. $\core \subset \g$\\
        \hline
        $N_r(\mathbf{s})$ & Neighbor node set of the node set $\mathbf{s}$. $r$ is a parameter. \\
        \hline
        $\mbox{coherence}(\core)$ & Coherence metric for a subgraph. \\
        \hline
        $\mbox{coherence}(u, v)$ & Coherence metric for an edge between $u$ and $v$.\\
        \hline
        $\mbox{contrast}(\g)$ & Contrast metric for a subgraph. \\
        \hline
        $\mbox{contrast}(u, v)$ & Contrast metric for an edge between $u$ and $v$.\\
        \hline
        $\mbox{penalty}(u)$ & Node penalty metric.\\
        \hline
        \end{tabular}
    }
    \vspace{-0.3cm}
    \end{table}

    \subsection{A General Family of Metrics}
        With a specific coherence metric, i.e., $\mbox{coherence}(\core)$, and a specific contrast metric, i.e., $\mbox{contrast}(\g)$, the contrast subgraph mining problem becomes the following two maximization problems.
        \begin{enumerate}[leftmargin=*,noitemsep]
            \item Identify the coherent core around the seed nodes:
            \begin{equation*}
                \hat{\core} = \argmax_{\seed \subset \core \subset N_r(\seed)} \mbox{coherence}(\core).
            \end{equation*}
            \item Induce the contrast subgraph from the identified coherent core:
            \begin{equation*}
                \hat{\g} = \argmax_{\hat{\core} \subset \g \subset N_r(\hat{\core})} \mbox{contrast}(\g).
            \end{equation*}
        \end{enumerate}

        It is natural to start from a single edge, since an edge is also a subgraph of two nodes.
        We denote the edge coherence as $\mbox{coherence}(u, v)$ and the edge contrast as $\mbox{contrast}(u, v)$.
        Inspired by the formulation of the maximum dense subgraphs~\cite{goldberg1984finding}, we design the coherence/contrast metric based on the edge coherence/contrast as follows:
        \begin{eqnarray}
            \mbox{coherence}(\core) = \dfrac{\sum_{u, v \in \core \wedge u < v} \mbox{coherence}(u, v)}{\sum_{u \in \core} \mbox{penalty}(u)}, \label{eq:coherence}\\
            \mbox{contrast}(\g) = \dfrac{\sum_{u, v \in g \wedge u < v} \mbox{contrast}(u, v)}{\sum_{u \in \g} \mbox{penalty}(u)} \label{eq:contrast},
        \end{eqnarray}
        where $\mbox{penalty}(u)$ is a function to control the subgraph size. 
        The numerators in Equation~\ref{eq:coherence} and Equation~\ref{eq:contrast} become bigger when the subgraph grows.
        The idea of introducing node penalty is to penalize these large subgraphs. 
        In fact, the form of these two definitions is a generalization of the density function in~\cite{goldberg1984finding}.

        \noindent\textbf{Remark.}
        (1) When there is no user-provided seed (i.e., $\seed = \emptyset$), we will identify the most coherent subgraph among all non-empty subgraphs, which is the most ``similar'' part between $G_A$ and $G_B$.
        (2) We can find a series of non-overlapping contrast subgraphs by iteratively removing the newly added nodes in the most contrasting subgraph, i.e., $\hat{\g} \setminus \hat{\core}$, until $\hat{\g} \setminus \hat{\core} = \emptyset$. Here $\hat{\g}$ is the most contrasting subgraph in the current iteration, while $\hat{\core}$ is the same coherent core in all iterations. 

    \subsection{Specific Metrics}\label{sec:function_choice}

        Our algorithm works with any (1) non-negative edge coherence $\mbox{coherence}(u, v)$; (2) non-negative edge contrast $\mbox{contrast}(u, v)$; and (3) positive node penalty $\mbox{penalty}(u)$.
        Therefore, there are countless ways to define these metrics. 
        We discuss the specific metrics utilized in both Section~\ref{sec:exp} and Section~\ref{sec:app} here.

        \smallsection{Edge Coherence}
        Considering the real-world graphs are usually sparse, the non-zero edges are more telling than the zero-weighted edges.
        For example, in co-authorship graphs, the observed co-authorship node pairs are more important compared to any two unrelated nodes.
        Additionally, as discussed before, the edge weights are mainly about the connection strengths, and their distributions are similar.
        So in our experiments, for a given edge, we adopt its smaller weight in two graphs to describe its edge coherence.
        Formally, 
        \begin{equation}
            \mbox{coherence}(u, v) = \min\{E_A(u,v),  E_B(u,v)\}.
        \end{equation}
        
        \smallsection{Edge Contrast}
        A good contrast metric must meet the following requirements.
        \begin{enumerate}[leftmargin=*,noitemsep]
            \item \textbf{Symmetric.} The order of $G_A$ and $G_B$ should have no effect on the contrast metric. Formally, given two nodes, $u$ and $v$, if we swap $E_A$ and $E_B$, $\mbox{contrast}(u, v)$ should be the same.
            \item \textbf{Zero.} It is natural to require $\mbox{contrast}(u, v) = 0$ when the edge weights in two graphs are the same. That is, when $E_A(u,v) = E_B(u,v)$, $\mbox{contrast}(u, v)$ must be $0$.
            \item \textbf{Monotonicity.} Suppose $E_A(u,v) \le E_B(u,v)$. If we increase $E_B(u,v)$ and keep the other same, the score $\mbox{contrast}(u, v)$ should increase. If we decrease $E_A(u,v)$ and keep the other same, the score $\mbox{contrast}(u, v)$ should also increase.
        \end{enumerate}

        Starting from these three requirements, we find a neat definition as follows.
        \begin{equation}
            \mbox{contrast}(u, v) = |E_A(u,v) - E_B(u,v)|
        \end{equation}
        It meets all the requirements and is therefore adopted in our experiments.

        \noindent\textbf{Node Penalty.}
        If there is no prior knowledge about the two graphs, a uniform node penalty is always a safe choice.
        Therefore, we use $\mbox{penalty}(u) = 1$ in all our experiments.


%% file: 2-related_work.tex

\section{Related Work}\label{sec:rel}

    In this section, we introduce the related work on dense graph patterns within a single graph, cross-graph patterns, and interesting phrase patterns of contrast in the text mining field.

    \subsection{Dense Subgraphs within a Single Graph}
        Our coherence and contrast metrics are inspired by the dense subgraph problem, i.e., finding a subset of nodes that maximizes the ratio of the number of edges between these nodes over the number of selected nodes.
        A polynomial time algorithm using a network flow model is first proposed to solve this problem~\cite{goldberg1984finding}.
        \cite{asahiro2000greedily} tries to solve it approximately using a greedy algorithm.
        Beyond traditional dense subgraphs, it has been proved that if one wants to specify the subgraph size, the problem becomes NP-hard~\cite{khot2006ruling}. Gibson et al.~\cite{gibson2005discovering} design an efficient algorithm specifically for giant dense subgraphs. Rossi et al.~\cite{rossi2014fast} develop a fast, parallel maximum clique algorithm for sparse graphs; and Leeuwen et al.~\cite{van2016subjective} explore the most interesting dense subgraphs based on the user's prior belief. 

    \subsection{Cross-Graph Patterns}

        \noindent\textbf{Frequent Subgraph.}
        Given a large collection of graphs, frequent subgraph mining aims to find frequent structures across different graphs~\cite{yan2002gspan,yan2003closegraph}.
        It is an unsupervised task but focuses on shared patterns instead of contrast patterns.
        Moreover, it usually requires more than two, usually tens of graphs as input.

        \noindent\textbf{Discriminative Subgraph.}
        In discriminative subgraph mining, given multiple graphs with their class labels, the goal is to figure out the subgraphs that discriminate between classes~\cite{thoma2010discriminative,ting2006mining,jin2011lts}.
        Discriminative subgraph is sometimes named as contrast subgraph~\cite{ting2006mining}, however, it's completely different from our contrast subgraph mining, due to its supervised learning nature.

        \noindent\textbf{Co-Dense Subgraph.}
        Density is an important measurement for graphs.
        Among various models and definitions of dense subgraphs, quasi-clique is one of the most prominent ones~\cite{liu2008effective}.
        As a natural extension, researchers have explored how to efficiently find cross-graph quasi-cliques~\cite{pei2005mining} and coherent subgraphs~\cite{boden2012mining,boden2017mimag}.
        Jiang et al. further extend cross-graph quasi-cliques to frequent cross-graph quasi-cliques~\cite{jiang2009mining}.
        As an application, coherent dense subgraph mining has shown its usefulness in biological networks~\cite{hu2005mining}.
        We can also generate some contrast subgraph in terms of density by straightforwardly applying co-dense subgraph mining methods on $G_A, \overline{G_B}$ and then on $\overline{G_A}, G_B$. 
        However, as shown in our experiments, these algorithms become extremely slow because the complementary graphs are so dense that pruning techniques are no longer effective.

    \subsection{Informative Phrases from the Contrast}
        Given two corpora (\ie, the main corpus and the background corpus), informative phrases are the phrases that occur more frequently in the main corpus than the background corpus~\cite{monroe2008fightin,bedathur2010interesting,gao2012top,majumdar2014fast,liu2015mining,shang2017automated}.
        With the concept of contrast, informative phrases are more salient than routine phrases like those consisted of stopwords, although their frequencies in the main corpus are not super high.
        The definition of informative phrases is based on individual phrase frequencies without considering a set of phrases as a whole. Such definition is not applicable for subgraphs.

%% file: 4-method.tex

\section{Methodology}\label{sec:method}

    In this section, we derive the polynomial-time algorithm for the contrast subgraph mining.
    Since the forms of coherence and contrast metrics are similar, we focus on extracting contrast subgraphs in the derivation.
    For identifying coherent cores, the same derivation applies if one replaces the edge contrast $\mbox{contrast}(u, v)$ with the edge coherence $\mbox{coherence}(u, v)$.

    \subsection{Binary Search}

        First of all, we reduce the original maximization problem to a verification problem by applying the binary search technique on the contrast score.
        In order to conduct the binary search, we need to figure out the lower/upper bounds of the subgraph contrast as well as the minimum contrast difference between two subgraphs.

        \smallsection{Lower Bound}
        In the worst case, there is only one different edge between $G_A$ and $G_B$ (assuming they are not exactly same). Let $\epsilon_{c} > 0$ be the minimum non-zero edge contrast. We have
            \begin{equation*}
                \mbox{contrast}(\g) \ge \dfrac{\epsilon_{c}}{\sum_{u \in V} \mbox{penalty(u)}}
            \end{equation*}

        \smallsection{Upper Bound}
        As an ideal case, all edges are self-loops of the same node in one graph, while that node has no self-loops in the other graph. Let $\epsilon_p > 0$ be the minimum node penalty. We have
            \begin{equation*}
                \mbox{contrast}(\g) \le \dfrac{\sum_{u,v\in V \wedge u<v} \mbox{contrast}(u, v)}{\epsilon_p}
            \end{equation*}

        \smallsection{Minimum Contrast Difference}
        Given any two subgraphs $\g_1, \g_2$, because the edge contrast difference is at least $\epsilon_c$ and the denominator is no more than the total node penalty $\sum_{u \in V} \mbox{penalty}(u)$, we have
        \begin{equation*}
            |\mbox{contrast}(\g_1) - \mbox{contrast}(\g_2)| > \dfrac{\epsilon_c}{(\sum_{u \in V} \mbox{penalty}(u))^2}
        \end{equation*}
        
        Then, the key problem becomes that given a binary searched contrast score $\delta$, whether we can find a subgraph $\g$ such that $\core \subset \g \subset N_r(\core)$ and $\mbox{contrast}(\g) \ge \delta$.
        Specifically, we need an algorithm that can efficiently check whether $\mbox{contrast}(\hat{\g}) \ge \delta$ for any $\delta \ge 0$.
        It is equivalent to checking the following inequalities.
\small
        \begin{eqnarray*}
                        & \max_{\core \subset \g \subset N_r(\core)} & \dfrac{\sum_{u, v \in \g \wedge u < v}\mbox{contrast}(u, v)}{\sum_{u \in \g} \mbox{penalty}(u)} \ge \delta \\
                        & & \mbox{\color{blue}{/* because penalties are positive */}} \\
        \Leftrightarrow & \max_{\core \subset \g \subset N_r(\core)} & \sum_{u, v \in \g \wedge u < v}\mbox{contrast}(u, v) - \sum_{u \in g} \delta \cdot \mbox{penalty}(u) \ge 0 \\
        \Leftrightarrow & \min_{\core \subset \g \subset N_r(\core)} & \sum_{u \in \g} \delta \cdot \mbox{penalty}(u) - \sum_{u, v \in \g \wedge u < v}\mbox{contrast}(u, v) \le 0 \\
                        & & \color{blue}{\mbox{/* let}\ d(u) = \sum_{v \in N_r(\core)} \mbox{contrast}(u, v)\mbox{ */}} \\
        \Leftrightarrow & \min_{\core \subset \g \subset N_r(\core)} & \sum_{u \in \g} \delta \cdot \mbox{penalty}(u) \\
                        &         & - \frac{1}{2}\Big(\sum_{u \in \g} d(u) - \sum_{u \in \g, v \in N_r(\core) \setminus \g} \mbox{contrast}(u, v)\Big) \le 0 \\
        \Leftrightarrow & \min_{\core \subset \g \subset N_r(\core)} & \sum_{u \in \g} \Big(2 \cdot \delta \cdot \mbox{penalty}(u) - d(u)\Big) \\
                        &         & + \sum_{u \in \g, v \in N_r(\core) \setminus \g} \mbox{contrast}(u, v) \le 0
        \end{eqnarray*}
\normalsize

        \begin{figure}[t]
          \centering
          \includegraphics[width=0.45\textwidth]{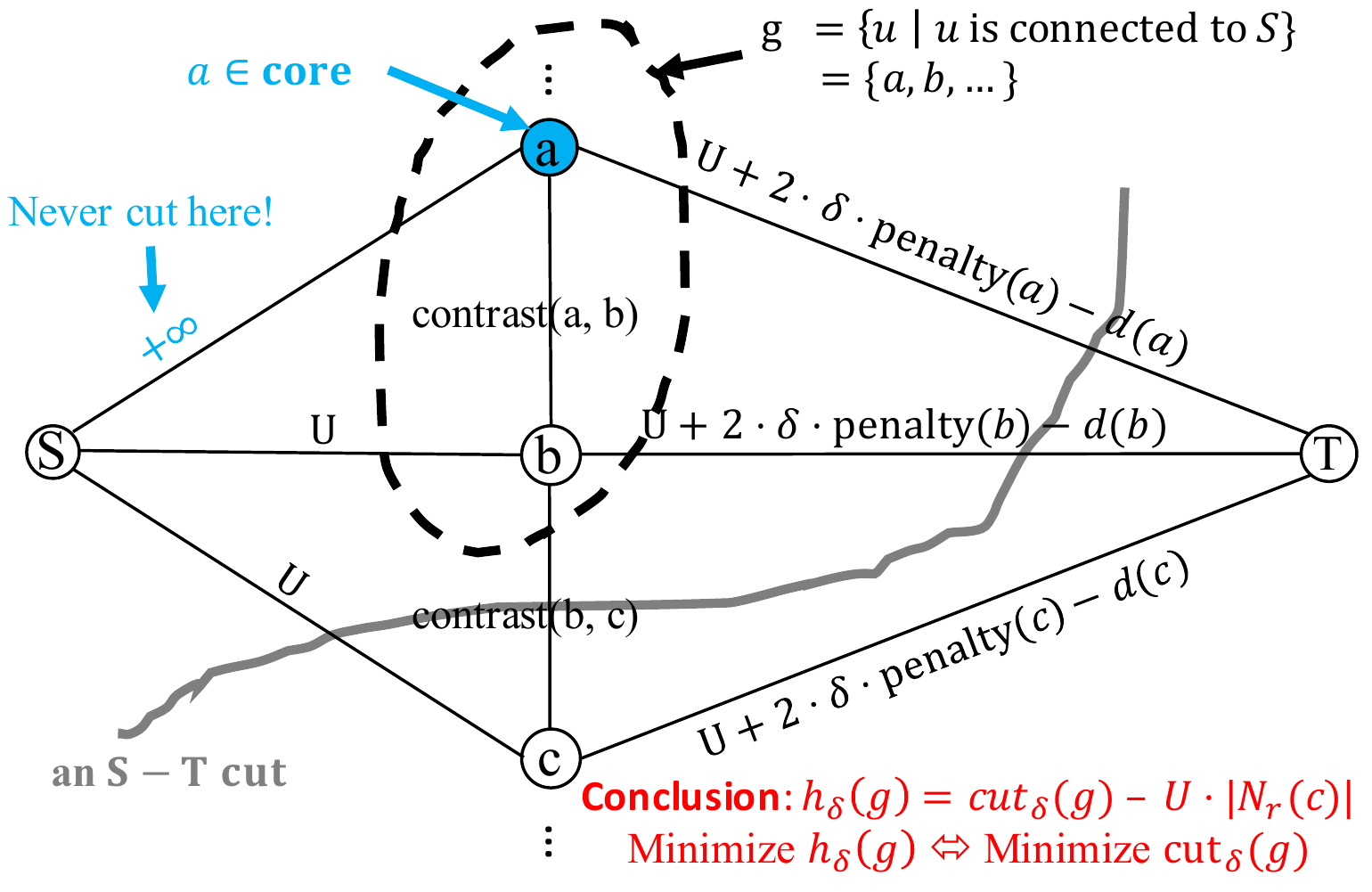}
          \vspace{-0.4cm}
          \caption{An Illustration for the Min-Cut Model.}\label{fig:network}
          \vspace{-0.4cm}
        \end{figure}

        Therefore, the problem is transformed to minimize the following function $h_\delta(\g)$ with the constraints $\core \subset \g \subset N_r(\core)$.
        \begin{equation*}
        h_\delta(\g) = \sum_{u \in \g} \Big( 2 \cdot \delta \cdot \mbox{penalty}(u) - d(u)\Big) + \sum_{u \in g, v \notin g} \mbox{contrast}(u, v)
        \end{equation*}

    \subsection{Min-Cut Derivation}\label{sec:mincut}

        Fortunately, we find the aforementioned minimization problem for the function $h_\delta(\g)$ is solvable in a polynomial time.

        We first construct a network with a source node $\mathcal{S}$ and a sink node $\mathcal{T}$ as shown in Figure~\ref{fig:network}, then show a one-to-one mapping between its $\mathcal{S}-\mathcal{T}$ finite-value cuts and the contrast subgraphs $\g$ meeting the constraints. Finally, we prove that the min cut gives us the most contrasting subgraph.

        \smallsection{Constructed Network} 
        The network is constructed as follows.
        \begin{itemize}[leftmargin=*,noitemsep]
        \item Creates a source node $\mathcal{S}$ and a sink node $\mathcal{T}$.
        \item $\forall u \in \core$, adds an edge between $\mathcal{S}$ and $u$ of a weight $+\infty$.
        \item Chooses a large enough constant $U$ to make sure that all following edge weights are positive. In our implementation, $U$ is set as $\sum_{u, v} contrast(u, v)$, which is no smaller than any $d(u)$.
        \item $\forall u \in N_r(\core) \setminus \core$, adds an edge between $\mathcal{S}$ and $u$ of a weight $U$.
        \item $\forall u \in N_r(\core)$, adds an edge between $u$ and $\mathcal{T}$ of a weight $U + 2 \cdot \delta \cdot \mbox{penalty}(u) - d(u) $.
        \item $\forall \mbox{contrast}(u, v) \neq 0$, adds an edge between $u$ and $v$ of a weight $\mbox{contrast}(u, v)$.
        \end{itemize}

        \noindent\textbf{One-to-One Mapping.} Here, we show the one-to-one mapping between its $\mathcal{S}-\mathcal{T}$ finite-value cuts and the contrast subgraphs $\g$ meeting the constraints.
        Any $\mathcal{S}-\mathcal{T}$ cut can be uniquely identified by the set of nodes connected to the source node $\mathcal{S}$, which is denoted as $s(\mathcal{S}-\mathcal{T})$.
        Note that, we have a hard constraint that $\core \subset \g$.
        Therefore, any valid $\g$ should contain all nodes in $\core$, which means there are only $2^{|N_r(\core)|-|\core|}$ valid $\g$'s.
        Discarding those cuts of $+\infty$ cost (i.e., including those edges from $\mathcal{S}$ to $\core$), we will also have $2^{|N_r(\core)| - |\core|}$ different cuts left.
        So mapping the remaining $\mathcal{S}-\mathcal{T}$ cuts to $s(\mathcal{S}-\mathcal{T})$ forms a one-to-one mapping between the cuts and contrast subgraphs.

        Moreover, the value of the $\mathcal{S}-\mathcal{T}$ cut can be transformed to the $h_\delta(\g)$ function.
        Suppose we have a $\mathcal{S}-\mathcal{T}$ cut.
        According to the one-to-one mapping, we know that $\g = s(\mathcal{S}-\mathcal{T})$.
        The cut edges can be grouped into three types: (1) The edges between $\g$ and $\mathcal{T}$; (2) The edges between $\mathcal{S}$ and $N_r(\core) \setminus \g$; And (3) the edges between $\g$ and $N_r(\core) \setminus \g$.
        The sum of these edge weights are
        \begin{eqnarray*}
            \mbox{cut}_\delta(\mathcal{S}-\mathcal{T}) & = & \sum_{u \in \g} (U + 2 \cdot \delta \cdot \mbox{penalty}(u) - d(u)) \\
                                                       &   & + \sum_{v \in N_r(\core) \setminus \g} U  + \sum_{u \in \g, v \in N_r(\core) \setminus \g} \mbox{contrast}(u, v) \\
                          & = & |N_r(\core)| \cdot U + \sum_{u \in \g} (2 \cdot \delta \cdot \mbox{penalty}(u) - d(u)) \\
                          &   & + \sum_{u \in \g, v \in N_r(\core) \setminus \g} \mbox{contrast}(u, v)
        \end{eqnarray*}
        Therefore, if we follow the one-to-one mapping, we have
        \begin{equation*}
            h_\delta(\g) = \mbox{cut}_\delta(\mathcal{S}-\mathcal{T}) = \mbox{cut}_\delta(\g) - |N_r(\core)| \cdot U
        \end{equation*}

        \noindent\textbf{Final Solution.} 
        As a result, minimizing $h_\delta(\g)$ is equivalent to find the min cut in the constructed network.
        Thanks to the duality between the min cut and the max flow, we can locate the min-cut among the exponential number of candidates within a polynomial time~\cite{papadimitriou1998combinatorial}.
        More specifically, treating the edge weights in the constructed network as capacities, we calculate the max flow from $\mathcal{S}$ to $\mathcal{T}$. In the residual network, the nodes connected to $\mathcal{S}$ form $\hat{\g}$. To obtain the result node set, we do a depth-first-search in the min-cut residual graph from $\mathcal{S}$ and return all reachable nodes. Algorithm~\ref{alg:contrast} presents the workflow. 

        About the constraints $\core \subset \g \subset N_r(\core)$, it is clear that the min-cut solution will only involve nodes in $N_r(\core)$ because they are the only nodes in the constructed network.
        Moreover, because the edge weights of all edges from $S$ to $\forall u \in \core$ are set to $+\infty$, they will never be included in the min cut.
        Therefore, the $\g$ induced from the min-cut will never miss any nodes in the coherent core $\core$.

    \SetAlgoSkip{}
    \begin{algorithm}[t]
        \caption{Contrast Subgraph Mining from Coherent Core}\label{alg:contrast}
        \textbf{Require}: $G_A, G_B$, the coherent core $\core$, and the parameter $r$. \\
        \textbf{Return}: The most contrasting subgraph $\hat{\g}$ ($\core \subset \hat{\g} \subset N_r(\core)$). \\
        Compute $N_r(\core)$ using a breadth-first search (BFS). \\
        $\epsilon_c \leftarrow \min_{u,v\in V \wedge u<v} \mbox{contrast}(u, v)$ \\
        $\epsilon_p \leftarrow \min_{u,v\in V \wedge u<v} \mbox{penalty}(u, v)$ \\
        $l \leftarrow \dfrac{\epsilon_c}{\sum_{u \in V} penalty(u)}$ \\
        $h \leftarrow \dfrac{\sum_{u, v \in V \wedge u<v} \mbox{contrast}(u, v)}{\epsilon_p}$ \\
        $\epsilon \leftarrow \dfrac{\epsilon_c}{\Big(\sum_{u \in V} \mbox{penalty}(u)\Big)^2}$ \\
        $\hat{\g} \leftarrow \core$ \\
        \While {$|h - l| > \epsilon $} {
            $m \leftarrow \dfrac{l + h}{2}$ \\
            Construct the network as described in Section~\ref{sec:mincut}. \\
            flow $\leftarrow$ the max flow of the constructed network \\
            \If {flow - $|N_r(\core)| \cdot U \le 0$} {
                $l \leftarrow m$ \\
                Update $\hat{\g}$ by the current $\mathcal{S}-\mathcal{T}$ cut.
            }
            \Else {
                $h \leftarrow m$ \\
            }
        }
        \Return $\hat{\g}$.
    \end{algorithm}

    \subsection{Time Complexity Analysis}

        We analyze the time complexity step by step.
        Given the parameter $r$ and seed nodes $\mathbf{s}$, we can obtain $N_r(\mathbf{s})$ in an $O(|V| + |E_A| + |E_B|)$ time through a breadth-first-search.
        Computing $\epsilon_c$ and $\epsilon_p$ costs $O(|V| + |E_A| + |E_B|)$ time.
        The binary search will have $\log_2 \frac{h - l}{\epsilon}$ iterations before it stops.
        The bottleneck inside the binary search is the max flow part. 
        The state-of-the-art max flow algorithm~\cite{orlin2013max} is $O(nm)$ by combining James B Orlin's algorithm (for the sparse network) and the KRT algorithm (for the dense network), where $n$ and $m$ are the numbers of nodes and edges in the constructed network.
        Therefore, the time complexity of the max flow part is $O\big(|N_r(\core)| (|N_r(\core)| + |E_A| + |E_B|)\big)$.
        So the overall time complexity becomes $O\big(|V| + |E_A| + |E_B| + |N_r(\core)| (|N_r(\core)| + |E_A| + |E_B|) \log_2 \frac{h - l}{\epsilon}\big)$.
        In the worst case, when $r$ is big enough, $|N_r(\core)|$ becomes $|V|$. 
        So the worst case time complexity is $O\big(|V|(|V| + |E_A| + |E_B|)\log_2 \frac{h - l}{\epsilon}\big)$
        Moreover, as both $\log_2(h - l)$ and $\log_2 {\epsilon}$ are no more than the input length, this time complexity is polynomial.

%% file: 5-experiment.tex

\section{Experiments}\label{sec:exp}

  In this section, we utilize the collaboration change detection task as a sanity check.
  We first present the useful results from contrast subgraph mining.
  Then, we discuss the importance of the coherent core and neighborhood constraints.
  In the end, we compare our algorithm with an adapted co-dense subgraph mining method, regarding both effectiveness and efficiency. 
  We implement both our algorithm and the compared method in C++. 
  The following execution time experiments were conducted on a machine with Intel(R) Xeon(R) CPU E5-2680 v2 @ 2.80GHz using a single thread.

  \subsection{Collaboration Change Detection}

    \smallsection{The Two Graphs}
    Two co-authorship graphs are constructed by splitting the DBLP publication network dataset\footnote{\url{https://aminer.org/citation}} into two time periods (2001--2008 and 2009--2016).     
    The nodes are authors who published at least one paper in KDD, ICDM, WWW, ICML, and NIPS.
    The co-authorship from 2001 to 2008 forms the edges in $G_A$, while $G_B$ consists of co-authorships from 2009 to 2016. 
    It ends up with $6,999$ nodes and $17,806$ edges in total.
    The edge weight is the log value (i.e., $\log(x) + 1$) of the actual collaboration time $x$.
    We choose $r = 1$.

    \smallsection{Visualization}
    We use two aligned heat maps (a black vertical line in between) to visualize adjacency matrices of the subgraphs.
    The darker color, the larger weight.
    The coherent core is always positioned in the bottom-right corner and enclosed by red squares, while the nodes are ranked by their total edge weights of $E_A$ within this subgraph. 
    Moreover, we append a table of the node names in the top-down order of the rows in the heat map.


    \smallsection{Expectations}
    In this task, coherent cores are expected to be some long-term collaborators of the seed author spanned over 2001 -- 2016, and contrast subgraphs will be likely the researchers who have more collaborations with the seed author in only one of the two periods.

\begin{figure}[t]
      \centering
      \includegraphics[width=0.5\textwidth]{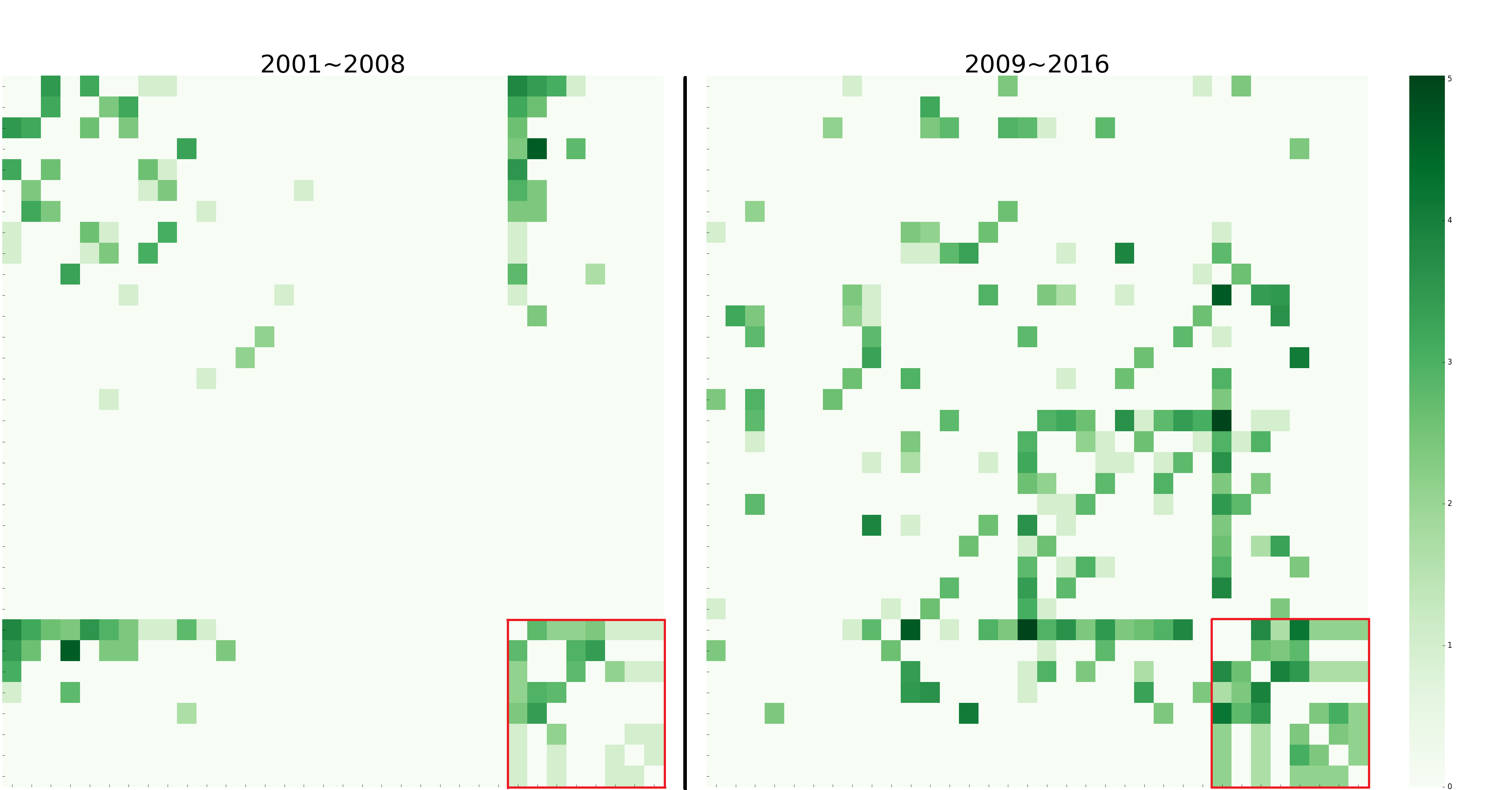}\\
    \scalebox{0.8}{
      \begin{tabular}{|c|l|}
      \hline
      $\seed$ & Jiawei Han \\
      \hline
      $\core$ & \multicolumn{1}{m{8.5cm}|}{Jiawei Han, Philip S. Yu, Jing Gao, Wei Fan, Charu C. Aggarwal, Latifur Khan, Bhavani M. Thuraisingham, Mohammad M. Masud} \\
      \hline
      $\g \setminus \core$ &  \multicolumn{1}{m{8.5cm}|}{Hong Cheng$^-$, Chen Chen$^-$, Xifeng Yan$^-$, Haixun Wang$^-$, Dong Xin$^-$, Chao Liu$^-$, Feida Zhu$^-$, Qiaozhu Mei$^-$, Chengxiang Zhai$^-$, Jian Pei$^-$, Bo Zhao$^-$, Hanghang Tong, Dan Roth$^+$, Thomas S. Huang$^+$, Cindy Xide Lin$^+$, David Lo$^+$, Chi Wang$^+$, Yizhou Sun$^+$, Fangbo Tao$^+$, Xiao Yu$^+$, Brandon Norick$^+$, Marina Danilevsky$^+$, Feng Liang$^+$, Jialu Liu$^+$, Ahmed El-Kishky$^+$, Jie Tang$^+$}\\
      \hline
      $\mbox{contrast}(\g)$ & $8.99$ \\
      \hline
      \end{tabular}
    }
      \vspace{-0.3cm}
      \caption{Collaboration Change Detection. The darker, the more collaborations. In $\g \setminus \core$, $^-$ means they have more collaborations with Prof. Jiawei Han in the first time period, while $^+$ means they have more collaborations with Prof. Jiawei Han in the second time period.}\label{fig:contrast_advisor_jiawei}
      \vspace{-0.3cm}
    \end{figure}

    \smallsection{Results}
    Since the time periods start from 2001, we choose a senior professor, Prof. Jiawei Han from the University of Illinois at Urbana-Champaign as the seed.
    As shown in Figure~\ref{fig:contrast_advisor_jiawei}, we find that all authors in the coherent core are long-term collaborators of Prof. Jiawei Han. 
    This contrast subgraph achieves a contrast score of $8.99$.
    Note that Jing Gao was Prof. Jiawei Han's Ph.D. student from 2006 to 2011, and 2008/2009 happens to be her midpoint. Moreover, they keep collaborating after Jing Gao's graduation.
    The first 11 researchers in $\g \setminus \core$ have much more collaborations with the researchers in the coherent core in 2001~--~2008 than those in 2009~--~2016. 
    The later 15 researchers have more collaborations from 2009 to 2016.
    In summary, the semantic meanings of these results match our observation and reality.
    Therefore, the contrast subgraph mining is helpful to the collaboration change detection task, and our algorithm is effective.


  \subsection{Coherent Core and Neighbor Constraints}
  \label{exp:coherent_core}

\begin{figure}[t]
      \centering
      \includegraphics[width=0.5\textwidth]{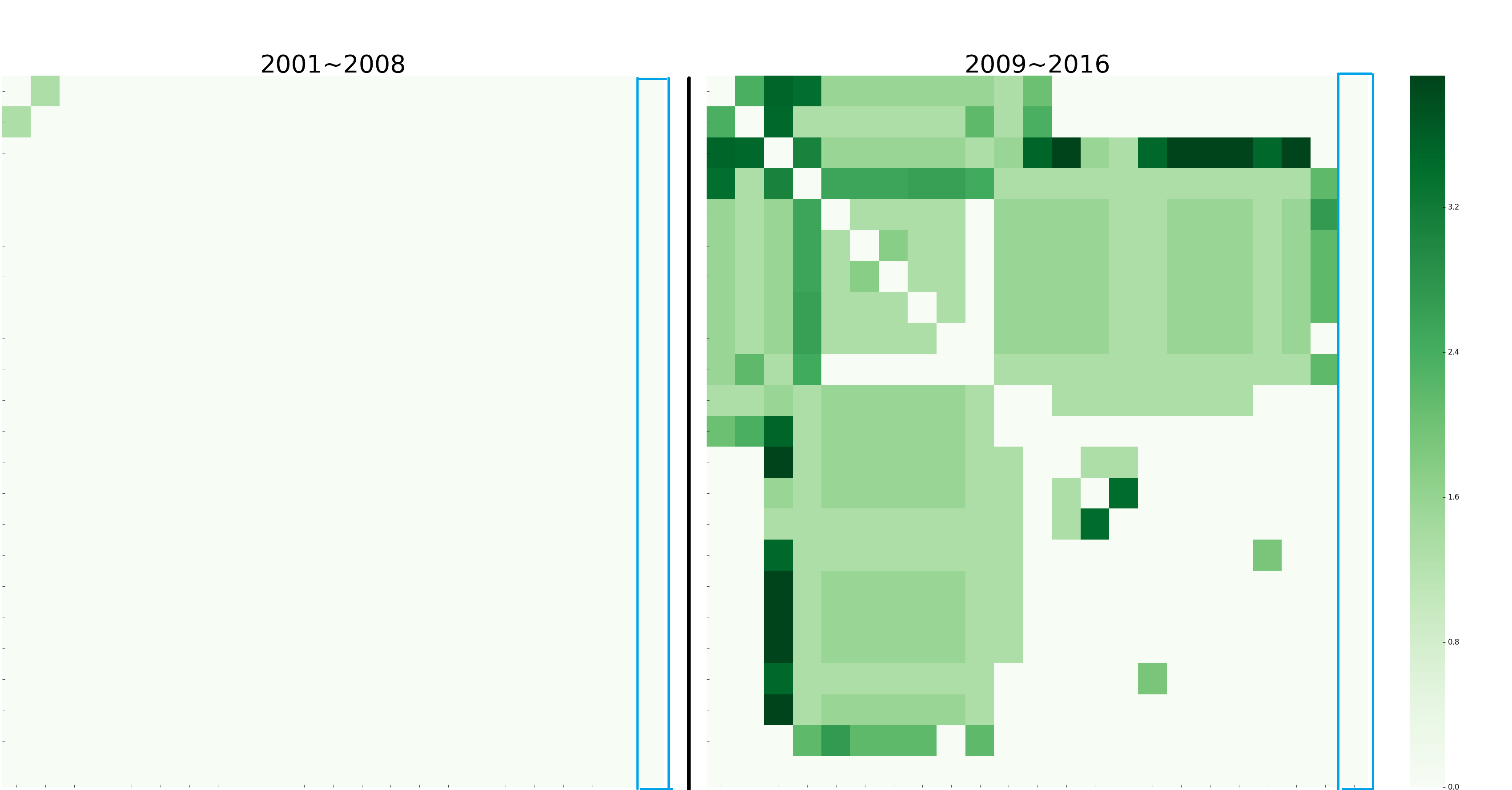}\\
    \scalebox{0.8}{
      \begin{tabular}{|c|l|}
      \hline
      $\seed$ & Jiwei Han \\
      \hline
      $\core$ & (no coherent core is computed) Jiawei Han \\
      \hline
      $\g \setminus \core$ &  \multicolumn{1}{m{8.5cm}|}{Chang-Tien Lu, Feng Chen, Achla Marathe, Naren Ramakrishnan, Nathan Self, Patrick Butler, Rupinder Paul Khandpur, Parang Saraf, Sathappan Muthiah, Wei Wang, Jose Cadena, Liang Zhao, Andy Doyle, Anil Vullikanti, Chris J. Kuhlman, Bert Huang, Jaime Arredondo, Graham Katz, David Mares, Lise Getoor, Kristen Maria Summers, Fang Jin}\\
      \hline
      $\mbox{contrast}(\g)$ & $11.08$ \\
      \hline
      \end{tabular}
    }
      \vspace{-0.3cm}
      \caption{[Without $\core$ and $N_r(\core)$] Collaboration Change Detection. 
      In the $\g \setminus \core$ part, none of them has collaborated with the seed.
      The results become meaningless.}\label{fig:contrast_no_core}
      \vspace{-0.3cm}
    \end{figure}

    \smallsection{If NOT inducing from coherent cores}
    If we directly extract the most contrasting subgraph without the coherent core step and the neighborhood step, we will find, of course, a subgraph with a potentially higher contrast score. 
    We run the collaboration change detection task using Prof. Jiawei Han as the seed again but skip these two steps.
    As shown in Figure~\ref{fig:contrast_no_core}, we obtain a subgraph of a contrast score $11.08$, which is higher than the previous score (i.e., $8.99$).
    Despite the high contrast, as highlighted by the light blue box in the figure, the seed node Prof. Jiawei Han never collaborated with anyone in the contrast subgraph.
    Therefore, although this subgraph has a high contrast score, we believe its semantic is unrelated to the expected collaboration change detection.
    This example shows the necessity of inducing from coherent cores and enforcing the neighbor constraint, which acts as anchors to make the induced contrast subgraph semantically stable and smooth.

\begin{figure}[t]
      \centering
      \includegraphics[width=0.5\textwidth]{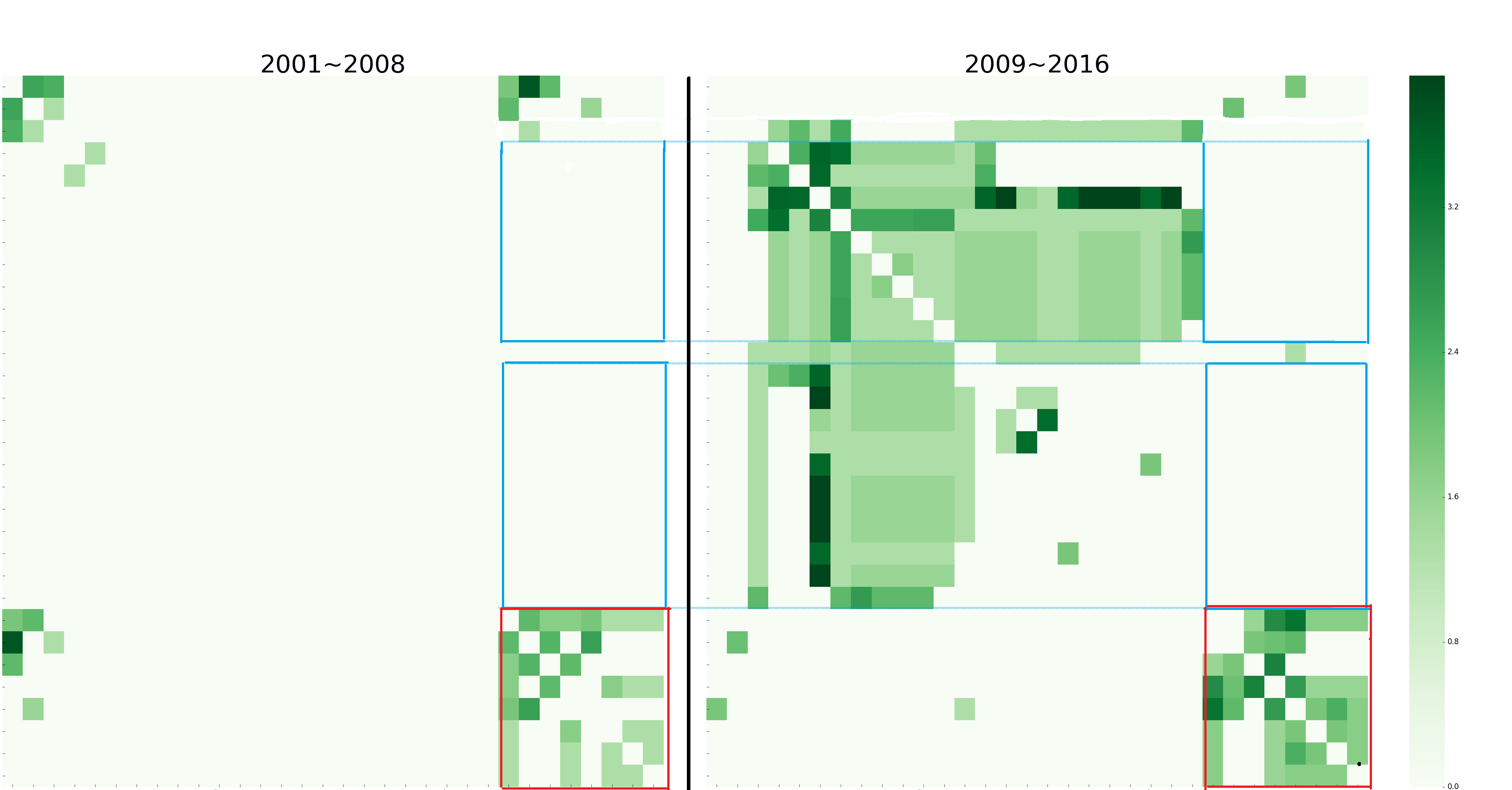}\\
    \scalebox{0.8}{
      \begin{tabular}{|c|l|}
      \hline
      $\seed$ & Jiawei Han \\
      \hline
      $\core$ & \multicolumn{1}{m{8.5cm}|}{Jiawei Han, Philip S. Yu, Wei Fan, Jing Gao, Charu C. Aggarwal, Latifur Khan, Bhavani M. Thuraisingham, Mohammad M. Masud} \\
      \hline
      $\g \setminus \core$ & \multicolumn{1}{m{8.5cm}|}{Haixun Wang, Jian Pei, Wei Wang, Chang-Tien Lu, Feng Chen, Achla Marathe, Naren Ramakrishnan, Nathan Self, Patrick Butler, Rupinder Paul Khandpur, Parang Saraf, Sathappan Muthiah, Jose Cadena, Liang Zhao, Andy Doyle, Anil Vullikanti, Chris J. Kuhlman, Bert Huang, Jaime Arredondo, Graham Katz, David Mares, Lise Getoor, Kristen Maria Summers, Fang Jin}\\
      \hline
      $\mbox{contrast}(\g)$ & $9.37$ \\
      \hline
      \end{tabular}
    }
      \vspace{-0.3cm}
      \caption{[Without $N_r(\core)$] Collaboration Change Detection. 
      Most researchers in $\g \setminus \core$ have no collaboration with the researchers in $\core$.
      The results are noisy and less meaningful.}\label{fig:contrast_no_grow}
      \vspace{-0.3cm}
    \end{figure}

    \smallsection{If NOT considering the $N_r(\core)$ constraint}
    Suppose we only utilize the coherent core, but we drop the $N_r(\core)$ constraint.
    Again, we can find a subgraph of a higher contrast score, as shown in Figure~\ref{fig:contrast_no_grow}. 
    Its contrast score is $9.37$, which is higher than the previous result, $8.99$.
    However, the results are noisy because of involving many researchers who never worked with any researcher in the coherent core.
    We cannot locate the researchers that are specifically related to the researchers in the core because there is no guarantee for the distance from the researchers in our coherent core.
    For example, as highlighted by two blue rectangles in the figure, those researchers have absolutely no connection to our coherent core.
    For sure, this result looks better than the one without any constraint, but it is still not as good as Figure~\ref{fig:contrast_advisor_jiawei}, in terms of semantic meanings.
    Therefore, the neighbor constraint is necessary to bridge the contrast subgraph to the coherent core and make them more semantically connected.

  \subsection{Running Time and Contrast Comparisons}

    To our best knowledge, we are the first to mine informative contrast subgraphs between two graphs. 
    There is few attempt before to attack this problem directly.
    In order to make comparisons, we adapt the co-dense subgraph mining algorithm~\cite{boden2012mining,boden2017mimag} by using $(G_A, \overline{G_B})$ and $(G_B, \overline{G_A})$ as inputs.
    We name it as \emph{adapted co-dense algorithm}.
    This algorithm first finds several subgraphs which are dense in one graph but sparse in the other graph.
    From these results, we then pick the subgraph of the highest $\mbox{contrast}(\g)$ score as the contrast subgraph.

\begin{figure}[t]
      \centering
      \subfigure[Running Time]{
        \includegraphics[width=0.22\textwidth]{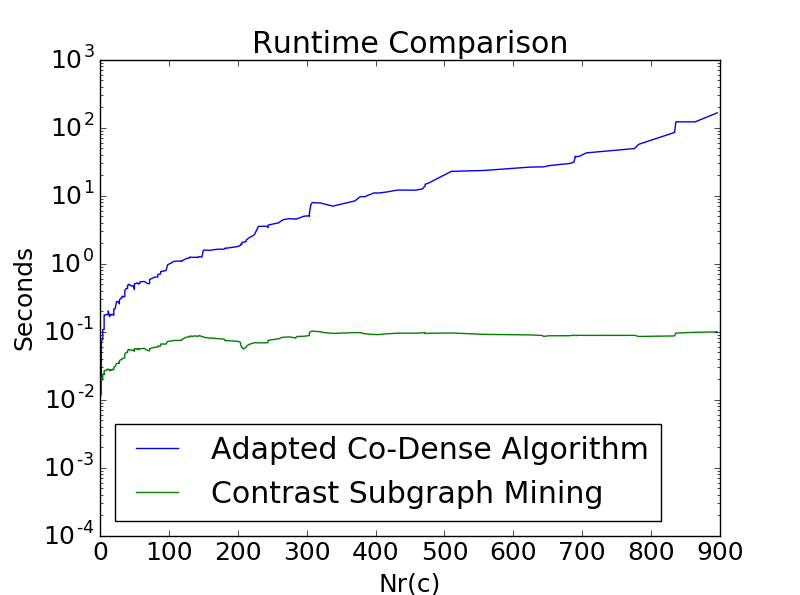}
        \label{fig:runtime_comparison}
      }
      \subfigure[Contrast Scores]{
        \includegraphics[width=0.22\textwidth]{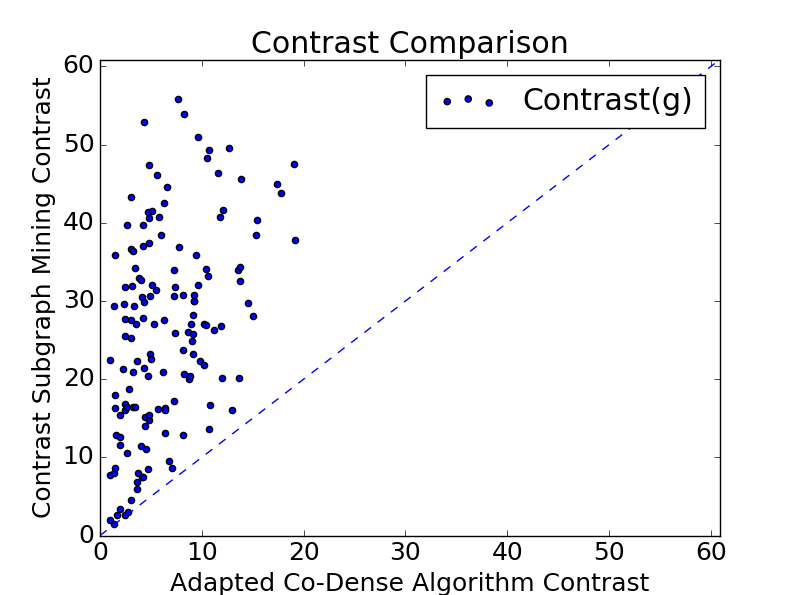}
        \label{fig:contrast_comparison}
      }
      \vspace{-0.4cm}
      \caption{Comparison between the adapted co-dense algorithm and our contrast subgraph mining algorithm.}
      \vspace{-0.4cm}
    \end{figure}

    We randomly pick a node as the seed and only pass the edges within $N_r(\seed)$ to the adapted co-dense algorithm. 
    Figure~\ref{fig:runtime_comparison} plots the running times of both methods after the same $\seed$ is given.
    Note that the Y-axis is log-scale.
    From the figure, one can observe that the adapted co-dense algorithm requires an exponential running time w.r.t. the number of neighbor nodes. 
    The primary reason is that the complementary graphs are so dense that most of, if not all, pruning techniques are no longer effective.
    Our algorithm demonstrates a polynomial growth as same as our previous time complexity analysis.

    We also compare the contrast scores achieved by the two algorithms, as shown in Figure~\ref{fig:contrast_comparison}. 
    In this figure, each data point corresponds to two runs of the same seed.
    For a given seed, we record the contrast score of the adapted co-dense algorithm as X and the contrast score of our contrast subgraph mining algorithm as Y.
    Therefore, the data points above $y=x$ indicates better performance than the adapted co-dense algorithm, vice versa.
    We plot the $y=x$ line for reference. 
    It is not surprising that the contrast subgraph mining algorithm will always have a higher score than the adapted co-dense algorithm because they have slightly different objectives.
    The ``contrast'' obtained from the adapted co-dense algorithm emphasizes more on the density vs. the sparsity.
    This is from a macro view, while our $\mbox{contrast}(\g)$ is defined in a micro view by considering every edge.

    In summary, our contrast subgraph mining algorithm significantly outperforms the adapted co-dense subgraph mining algorithm regarding both efficiency and effectiveness.

\section{Real-World Applications} \label{sec:app}
  In this section, we present two real-world applications of contrast subgraph mining: (1) spatio-temporal event detection based on a taxi trajectory dataset and (2) trend detection on e-commerce platforms based on an Amazon rating dataset. 

  \subsection{Spatio-Temporal Event Detection}

    \smallsection{The Two Graphs}
    We construct two graphs using the traffic data and road network in Beijing.
    The same taxi trajectory dataset in~\cite{hong2015detecting} is utilized.
    We use the road network data of Beijing, which contains 148,110 road joints as nodes and 96,307 road segments as edges.
    The GPS trajectories were generated by 33,000 taxis during a period of 30 days in November 2012.
    Taxi trajectories are mapped onto the road network using map matching algorithm proposed in~\cite{yuan2010interactive}.
    For each road segment, its weight in $G_A$ is set as the number of taxis traveled along this road segment between 10:30 PM and 11:00 PM on November 24th, 2012, while its weight in $G_B$ is set as the average number of taxis traveled along this road segment between 10:30 PM and 11:00 PM from November 1st to November 23rd.
    The neighborhood parameter $r$ is set as $20$ considering the roads are segmented in a fine-grained way.

    \smallsection{Visualization}
    After identified the coherent core and extracted the contrast subgraph, we project them on a map and utilize a red dot, a blue circle, and a black dashed polygon to highlight the seed, the coherent core, and the contrast subgraph, respectively. 


    \smallsection{Expectations}
    In this task, coherent cores are expected to contain those road segments that are busy in both the real-time graph and the historical graph, and contrast subgraphs will be likely those road segments that are only busy in either the real-time graph of the historical graph. 

    \smallsection{Results}
    We choose some seeds on the East 2nd Ring, a major road in Beijing.
    As shown in Figure~\ref{fig:contrast_concert}, our algorithm first identifies a coherent core about the usual, busy traffic on the East 2nd Ring, and later extracts a contrast subgraph including many nodes and road segments around the Beijing Workers' Sports Complex.
    The contrast score is $23.87$.
    This contrast subgraph demonstrates a significantly busier real-time traffic than usual, therefore, it may reflect some unusual event.
    After some search on the Internet, interestingly, on November 24, 2012, there was a concert hosted at the Beijing Workers' Sports Complex, whose ending time was just around 10:30 PM.
    This finding further consolidates the usefulness of contrast subgraph mining and the effectiveness of our algorithm in real world applications.

  \subsection{Trend Detection on E-commerce Platform}

  \begin{figure}[t]
      \centering
      \includegraphics[width=0.45\textwidth]{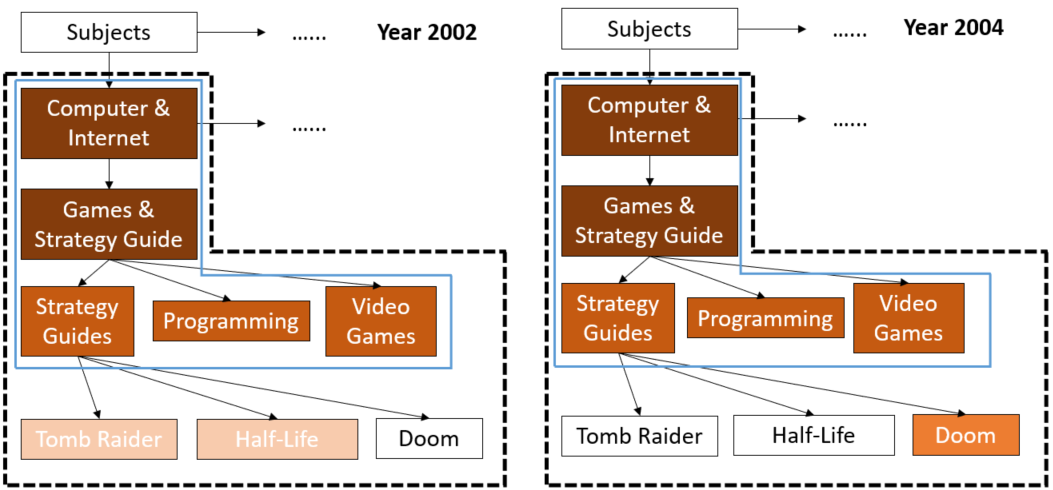}
    \scalebox{0.8}{
      \begin{tabular}{|c|l|}
      \hline
      $\seed$ & Games $\&$ Strategy Guides \\
      \hline
      $\core$ & \multicolumn{1}{m{8.5cm}|}{Computers $\&$ Internet, Games $\&$ Strategy Guides, Strategy Guide, Programming, Video Games} \\
      \hline
      $\g \setminus \core$ &  \multicolumn{1}{m{8.5cm}|}{Tomb Raider$^-$, Half-Life$^-$, Doom$^+$}\\
      \hline
      $\mbox{contrast}(\g)$ & $1.44$ \\
      \hline
      \end{tabular}
    }
      \vspace{-0.3cm}
      \caption{Trend Detection on E-commerce Platform. The darker, the more popular. In $\g \setminus \core$, $^-$ means more popular in 2002, while $^+$ means more popular in 2004.}\label{fig:ecommerce}
      \vspace{-0.3cm}
  \end{figure}

    \smallsection{The Two Graphs}
    We construct two trees based on Amazon's product hierarchy.
    The Amazon rating dataset~\cite{jindal2008opinion} is adopted to extract the tree and assign weights to each tree edge.
    There are 14,222 nodes in total.
    The root node is ``Subjects''.
    A node is represented as its prefix path from the root, such as ``Subjects$\rightarrow$Computers $\&$ Internet$\rightarrow$Games $\&$ Strategy Guides''.
    The weight of an edge linking from node $u$ to node $v$ equals to the log value (i.e., $\log(x) + 1$) of the total number of reviews $x$ received by the products under the node $v$.
    In $G_A$, the total number of received reviews are calculated based on the year 2002, while those in $G_B$ come from the year 2004.
    We set $r = 1$.
    
    \smallsection{Visualization}
    Considering these two graphs are trees, we directly visualize the tree structures.
    Only nodes within the contrast subgraphs are colored.
    The darker, the bigger weights of the edge from $v$'s parent node to $v$. 
    The blue polygon shows the coherent core and the dashed black polygon encloses the contrast subgraph.


    \smallsection{Expectations}
    In this task, we expect the nodes in the coherent core represent popular products types in both years, while the nodes in the contrast subgraph demonstrate the specific trends in one of the two years.

    \smallsection{Results}
    We choose ``Subjects$\rightarrow$Computers $\&$ Internet$\rightarrow$Games $\&$ Strategy Guides'' as the seed. 
    As shown in Figure~\ref{fig:ecommerce}, because the popularity of each sub-category under ``Games $\&$ Strategy Guides'' is similar in both years, they form the coherent core.
    Moreover, ``Tomb Raider'' and ``Half-Life'' received many reviews in 2002 but not in 2004, while ``Doom'' became popular in term of the review amount in 2004.
    Therefore, these three nodes become the contrast subgraph of a contrast score $1.44$.
    After some brief research in Wikipedia, we found out that ``Tomb Raider'' released ``The Prophecy'' episode in 2002 but nothing in 2004, while ``Doom'' released ``Doom 3'' in 2004 and its predecessor (i.e., ``Doom 64'') was long time ago in 1997. 
    These results are constructive for the sales trend analysis on e-commerce platforms and further prove the effectiveness of our algorithm on graphs of abstract nodes like hierarchical trees.

%% file: 6-conclusion.tex

\section{Conclusion \& Future Work}\label{sec:con}

In this paper, we formulated the important contrast subgraph mining problem. 
To avoid meaningless contrast subgraphs, we proposed to first identify coherent cores as cornerstones.
Our framework can admit a general family of coherence, contrast, and node penalty metrics.
After rigorous derivations, we developed an elegant polynomial-time algorithm to find the global optimum for this problem.
Extensive experiments verified the necessity of introducing coherent cores as well as the efficiency and effectiveness of our proposed algorithm.
Real-world applications demonstrated the tremendous potentials of the contrast subgraph mining.

In future, we will (1) extend the problem setting to more than two graphs; (2) find the top-$k$ contrast subgraphs with overlaps; and (3) apply our contrast subgraph mining algorithm to other tasks, like the abnormality detection in social networks.